\documentclass[11pt]{article}
\pdfoutput=1
\usepackage{jheppub}
\usepackage{graphicx,float,flowchart}
\usepackage{amssymb}
\usepackage{subcaption}
\usepackage{amsmath}
\usepackage{slashed}
\usepackage{hyperref}
\usepackage{caption}
\usepackage{xcolor}
\usepackage{dsfont}
\usepackage{verbatim}
\usepackage{setspace}
\usepackage{slashed}
\usepackage{xcolor}

\newcommand\beq{\begin{equation}}
\newcommand\eeq{\end{equation}}

\def\shrug{\texttt{\raisebox{0.75em}{\char`\_}\char`\\\char`\_\kern-0.5ex(\kern-0.25ex\raisebox{0.25ex}{\rotatebox{45}{\raisebox{-.75ex}"\kern-1.5ex\rotatebox{-90})}}\kern-0.5ex)\kern-0.5ex\char`\_/\raisebox{0.75em}{\char`\_}}}

\title{$T\bar{T}$ Deformation and the Complexity=Volume Conjecture}
\author{Hao Geng}
\affiliation{Department of Physics, University of Washington, Seattle, WA, 98195-1560, USA}
\emailAdd{hg666@uw.edu}
\preprint{\today}
\abstract{Complexity in quantum physics measures how difficult a state can be reached from a reference state and more precisely it is the number of fundamental unitary gates we have to operate to transform the reference state to the state we are considering. In the holographic context, based on several explicit calculations and arguments, it is conjectured that certain bulk volume calculates the boundary field theory subregion complexity. In this paper, we will show that the $T\bar{T}$ deformation shows a strong signal of the correctness of this complexity equals volume conjecture. A bonus is a way to look at the $T\bar{T}$ deformation, by its reversibility, as operating a unitary quantum circuit which prepares states in quantum field theory.}
\begin{document}
\maketitle
\section{Introduction}
Complexity measures how difficult it would be to accomplish a task. In the same spirit, in quantum physics, complexity of a quantum state $\rho$ measures how difficult it would be to prepare such a state. More concretely, it is quantified by the minimum amount of fundamental unitary gates one needs to operate to reach $\rho$ from a reference state. However, this suggestive definition involves several uncanonical jargons, for example what are the fundamental unitary gates and the reference state? These obstacles make it difficult to generalize the concept of complexity to a continuum quantum field theory without artificial and hence uncanonical identifications of the reference state, fundamental unitary gates and the measure of minimization. For example, such as those in the works \cite{Jefferson:2017sdb,Yang:2017nfn,Chapman:2017rqy,Hashimoto:2017fga}. The problems of those proposals are similar to using inelegant design scheme, imperfect logical gates and unnatural coding system to build a computer. Logically, for continuous systems like a quantum field theory on a generic curved spacetime background the circuit should reflect the fact that the system is continuous for example the existence of the energy-momentum tensor.

In the literature of black hole physics and anti-de-Sitter (AdS) holography, people conjectured that certain bulk volume is the complexity of the boundary quantum field theory up to some constants \cite{Alishahiha:2015rta,Susskind:2014rva,Stanford:2014jda}, which is the so-called Complexity=Volume (CV) proposal. From the bulk point of view, this is a pure geometric proposal which is easy to calculate. However, in this proposal the curvature scale of the bulk spacetime, which is not a fundamental constant, appears and the bulk volume is not a very natural concept from the boundary field theory point of view. More than this, CV does not capture the bulk physics beyond its geometry. Thereafter, people proposed the so called Complexity=Action (CA) conjecture \cite{Brown:2015bva,Brown:2015lvg} to avoid these issues. Using this CA proposal a bonus that AdS black holes saturate the famous Lloyh bound \cite{Lloyd_2000}, which says that the number of operations a computer can perform in a second is bounded by its energy, was uncovered in \cite{Brown:2015bva} which suggests that black holes might be the fastest computers. Nevertheless, the authors in \cite{Lehner:2016vdi} found that CA has a similar ambiguity as CV that there is a dependence on an unfundamental parameter when there is a null segment on the boundary of the bulk patch contributing to the action. Now it seems that CA and CV are of equally embarrassing footing as conjectures calculating the complexity of the boundary quantum field theory. Therefore more has to be said from the field theory point of view. 

In this paper, we will show that for a holographic field theory, described by the AdS/CFT correspondence \cite{Maldacena:1997re}, CV indeed calculates the complexity and point out several interesting questions to study and understand in future. Our tool is the holographic picture of the so called $T\bar{T}$ deformation \cite{Cavaglia:2016oda,Smirnov:2016lqw,McGough:2016lol}. We will briefly review essential aspects of $T\bar{T}$ for our study in Sec.\ref{sec:TT}. Then in Sec.\ref{sec:pic} we will discuss our main observations and argue that $T\bar{T}$ should be understood as operating a canonical unitary quantum circuit and hence calculating complexity in a natural way. In Sec.\ref{sec:EGEO} we will comment on how to understand the emergence of bulk geometry in our picture. Finally in Sec.\ref{sec:end} we will summarize our paper and point out interesting questions that can and should be studied in the soon future.
\section{Aspects of the $T\bar{T}$ Deformation}
\label{sec:TT}
In this section we will review several essential facts about the $T\bar{T}$ deformation for our study of complexity in the next section. A recent nice review for $T\bar{T}$ and $T\bar{T}$-like deformations is \cite{Jiang:2019hxb}. We will consider a d-dimensional field theory living on a curved spacetime with metric $h_{\alpha\beta}$.
\subsection{Field theory definition}\label{sec:def}
$T\bar{T}$ deformation for a quantum field theory is an explicit deformation of its Lagrangian density:
\begin{equation}
    \delta S=\delta\mu\int d^{d}x\sqrt{-h} T\bar{T}(x)
\end{equation}
where the operator $T\bar{T}(x)$ is a shorthand for the determinant of the energy-momentum tensor and $h$ is the determinant of the metric. Interesting properties of this deformation is that it preserves the solubility of the field theory \cite{Cavaglia:2016oda,McGough:2016lol,Bonelli:2018kik}, produces nonlocal effects \cite{Cardy:2019qao,Lewkowycz:2019xse}\footnote{A recent paper studying correlation functions in $T\bar{T}$ deformed CFTs (near IR) perturbatively is \cite{He:2019vzf}.} and more importantly is reversible. The reversibility is one of our motivations to identify the $T\bar{T}$ deformation as a unitary transformation or operating a quantum circuit. Moreover, this observation is very interesting if we consider the subregion complexity of the boundary field theory as we will discuss in detail in Sec.\ref{sec:pic} and Sec.\ref{sec:EGEO}. The picture is that we construct the subregion state $\rho_{R}$ starting from a single qubit and adding in more and more qubits along the way we are operating the circuit or the complementary picture that the circuit has multiple layers and the number of nontrivial gates on each layer is growing, from one two-qubit gate on the first layer, as we work through the circuit layer by layer. And in this complementary picture the input of the circuit is a tensor product state in a many-body Hilbert space (no entanglement among the single bodies making up the many-body state). 
\subsection{Holographic description}
From the point of view of the AdS/CFT correspondence  \cite{Maldacena:1997re}, the $T\bar{T}$ deformation of the conformal field theory living on the bulk UV slice is proposed to correspond to cut off the bulk at a certain radial slice $r_{c}$ in \cite{McGough:2016lol}. In this cutoff spacetime, the field theory dual is living on the cutoff slice $r=r_{c}$. Following \cite{McGough:2016lol} we write the cutoff AdS metric in the Fefferman-Graham form
\begin{equation}
    ds^{2}=\frac{dr^{2}}{r^{2}}+r^{2}g_{\alpha\beta}dx^{\alpha}dx^{\beta}, r\in[1,r_{c}]\label{eq:metric}.
\end{equation}
The dictionary between the $T\bar{T}$ deformation parameter $\mu$ and the bulk radial cutoff $r_{c}$ is
\begin{equation}
    \mu=\frac{16\pi G}{r_{c}^{2}}.\label{eq:dic}
\end{equation}
\section{$T\bar{T}$ implies Complexity=Volume}
\label{sec:pic}
In this section, we will explain our realization that $T\bar{T}$ deformation operates the canonical quantum circuit preparing the quantum field theory state and that circuit calculates the complexity. Our picture is motivated by AdS holography and we hope that it is generally true for any quantum field theory. We want to emphasize that we believe that the $T\bar{T}$ deformation realized as operating a quantum circuit is the canonical way to prepare the state and hence calculating complexity for continuous systems like quantum field theory. Because it uses the specific fact that for continuous systems we have the stress-energy tensor.

For the sake of the convenience of presentation, we will consider the bulk $AdS_{3}$ in global coordinate which can be put into the form \eqref{eq:metric}. In global coordinate, $AdS_{3}$ is a cylinder with its interior as the bulk and the time direction is pointing up along the cylinder. We will focus on a chosen time slice i.e. a cross section of the cylinder.

\begin{figure}
    \centering
    \includegraphics[width=6cm]{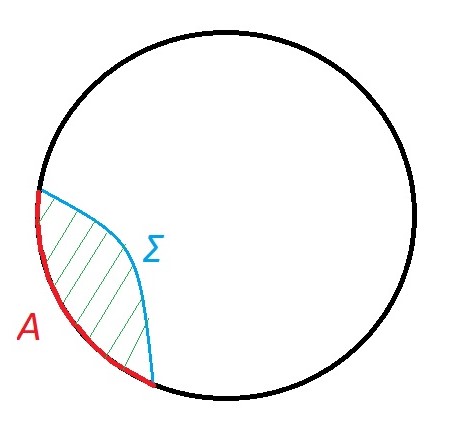}
        \caption{A time slice of $AdS_{3}$. The shaded region is the entanglement wedge of boundary subsystem A and the blue curve $\Sigma$ is its RT surface \cite{Ryu:2006bv}.}\label{pic:cplx}
\end{figure}

For a boundary subregion A, CV says that the complexity of its state $\rho_{A}$ is proportional to the volume of the entanglement wedge of A see \cite{Alishahiha:2015rta} and Fig.\ref{pic:cplx}.
We can see that this proposal is indeed in accord with the definition of complexity as the minimum amount of fundamental gates we have to use to construct a circuit that can prepare this state using $T\bar{T}$ deformation as follows. We start from a single qubit located at the point P for the field theory dual of the bulk region with the gold slice as the cutoff. As we do the opposite $T\bar{T}$ deformation more and more bulk regions are emerging and the field theory dual is living on outer and outer radial slices until we reach the boundary. To construct the subsystem A, we start from a single qubit located at P and along the way approaching to A we add in more and more qubits and so the subsystem size is growing or A is being constructed. But the point is that each time we add in more qubits we do a transformation of the whole already constructed subsystem (that inside the entanglement wedge of A) and the newly added qubits to make sure that in this bigger subsystem each qubit is correctly entangled with the others. Hence the whole circuit preparing A is in a layered structure that each layer has the amount of gates proportional to the size of the already constructed subsystem so far and so forth.

\begin{figure}
    \centering
    \includegraphics[width=6cm]{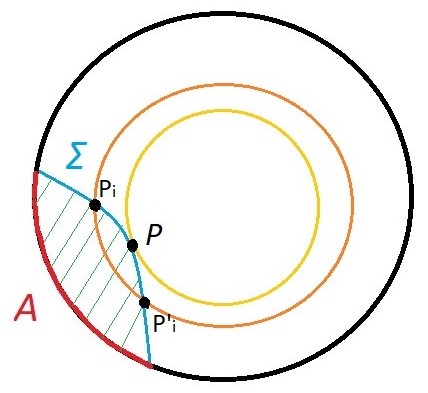}
        \caption{We do $T\bar{T}$ deformation the opposite way starting from IR to UV. For the subsystem A we start from the gold slice tangent to the RT surface $\Sigma$ at point P and opposite $T\bar{T}$ deformation constructs the subsystem A from P. An intermediate slice along the way is shown in orange which intersects $\Sigma$ at $P_{i}$ and $P'_{i}$.}\label{pic:CV}
\end{figure}

An example is shown in orange in Fig.\ref{pic:CV}. The subsystem constructed at that moment is $P_{i}P'_{i}$. Going one more step towards A by an infinitesimal opposite $T\bar{T}$ deformation is equivalent to adding some small amount of qubits and operating a circuit of size $P_{i}P'_{i}$ to make sure that the qubits on $P_{i}P'_{i}$ and the just added qubits are entangled correctly. This means that the entanglement structure of the qubits on $P_{i}P'_{i}$ is preserved when we go one step closer to A and so we only have to operate a circuit which entangles the just added qubits with the qubits on $P_{i}P'_{i}$ which implies that the circuit for this purpose is of size $P_{i}P'_{i}$. This says that each step is optimized and therefore this procedure is indeed calculating the complexity by using a minimum amount of gates.
As a result, the total amount of gates used or the complexity of the subregion state $\rho_{A}$ is equal to the sum of the amount of gates for each intermediate circuit or each layer of the whole circuit. In the continuous Language, it can be written as an integral:
\begin{equation}
    C(\rho_{A})=\sum_{n}(\text{\#of gates})_{n}=\int\delta n\times(\text{\# of gates})(n)
\end{equation}
where n is the step number and is parametrized by the $T\bar{T}$ deformation parameter $\mu$. Using the conclusion of our discussion that the optimized procedure is that for each step the number of gates should be proportional to the number of qubits in the already constructed subsystem, we have
\begin{equation}
    C(\rho_{A})\propto\int d[\mu] Area(\mu)
\end{equation}
where $Area(\mu)$ is the size of the already prepared subsystem at that step, for example $P_{i}P'_{i}$ in Fig.\ref{pic:CV} and $d[\mu]$ is a measure we do not see how to settle down. But what we do know is that $d[\mu]$ is the number of steps or the number of layers of the whole circuit that we have to go through to achieve an infinitesimal $T\bar{T}$ deformation $\delta\mu$. Now using our knowledge \eqref{eq:dic} we can write
\begin{equation}
    C(\rho_{A})\propto \int d[r]\times Area(r),
\end{equation}
so for $AdS_{d+1}$ if we have $d[r]=r^{d-1}dr$ then we get the CV proposal!

From the field theory point of view, for a 2d CFT, the measure $d[\mu]$ is given by
\begin{equation}
    d[\mu]=\frac{12\pi}{c\mu^{2}}d\mu\label{eq:measure}
\end{equation}
where we used the Brown-Henneaux central charge \cite{Brown:1986nw} $c=\frac{3}{2G}$ (notice as already happened in \eqref{eq:dic} we have set the bulk curvature radius to be one). Beyond this we do not have much to say, for now, on the field theory side, for example how to understand this layer density measure for the quantum circuit. One potential puzzle is the anti-proportionality to the central charge. Central charge measures the number of field contents for a CFT and so the larger the central charge is the more difficult it should be to prepare a state or at least it should not be easier. The answer is that in the whole expression for complexity we have to divide the volume by the Newton's constant G and hence multiply the volume by the central charge in the field theory language. This will cancel that central charge in the denominator of \eqref{eq:measure}.
\section{A Complementary Picture and the Emergence of Bulk Geometry}\label{sec:EGEO}
The previous section focuses on providing a tool to calculate complexity for field theory. Hence the interpretation is such that at each step there is a concrete field theory support living on a radial slice. From that point of view, the bulk geometry and hence $Area(\mu)$ and $d[\mu]$ are determined by the dual field theory state. These functions are hard to find without holography and hence holography helps us define the complexity for quantum field theories.

Now we provide a complementary picture of how $T\bar{T}$ works to the picture provided in the previous section. This picture does not need a concrete field theory support at each step. It only requires a many-body Hilbert space and a quantum circuit operated by the $T\bar{T}$ deformation. This supports the long-standing idea that geometry is an emergent concept. For the sake of convenience, we will assume that the fundamental quantum gates are two-body quantum gates.

The picture is motivated by the observation that subsystem complexity measures how much entanglement is there among the underlying degrees of freedom inside that subsystem or among the single-body states making up the many-body state describing that subsystem. It should not care about how much entanglement is there between the subsystem we are considering and its complement. Now instead of saying that we construct the subsystem staring from a single qubit located at P in Fig.\ref{pic:CV}, we interpret that the point P represents a tensor product state\footnote{As it is pointed out in \cite{Geng:2019ruz}, this is precisely one important starting point in the surface/state correspondence \cite{Miyaji:2015yva}.} in the many-body Hilbert space which does not have any entanglement among itself, for example saying that it is not a Bell state. After this we pretty much do the same as in the previous section. We input this state into the quantum circuit which has a layered structure and is operated by the $T\bar{T}$ deformation. At the very first step or layer there is only one two-body quantum gate and then as we go deeper and deeper into the circuit the number of two-body quantum gates is larger and larger at each layer. At each layer we expand the number of mutually entangled single-body states by operating a circuit with a bigger amount of two-body quantum gates than its previous step or layer. Finally we reach the state we want to prepare and that state contains an enough amount of information describing how the single-body states are entangled with each other.

In this picture, the bulk geometry is interpreted as the object or the structure encoding the history of how the state is canonically prepared and it is isomorphic to the whole circuit (a concrete realization of this picture in 1+1 dimensions has been made in \cite{Caputa:2017urj,Caputa:2017yrh}\footnote{We are very grateful for an anonymous person telling us the works \cite{Caputa:2017urj,Caputa:2017yrh}.}). This amount of information can also be recovered once we know every detail (how the single-body states making up the many-body state are entangled \footnote{It is worth to mention that this observation has been confirmed using the surface/state correspondence \cite{Miyaji:2015yva} in (pure) de-Sitter holography in \cite{Geng:2019ruz}. In that situation, for a boundary subsystem with size smaller than half of the whole system, the density matrix is a tensor product state and the entanglement wedge is empty.}) of the prepared many-body state. Hence we have the duality between gravity (geometry) and a boundary field theory and both of them have the same many-body Hilbert space. 

\section{Conclusions and Future Remarks}
\label{sec:end}
Motivated by holography and the $T\bar{T}$ deformation, in this paper we show that $T\bar{T}$ deformation could be understood as the canonical approach to prepare quantum states in a many-body quantum field theory and hence calculating complexity in quantum field theory. Along the way, we show that the Complexity=Volume proposal is indeed in accord with the definition of complexity. The reversibility of the $T\bar{T}$ deformation is the key element in this observation. Since our work only involves field theory language, it can potential be applied to general quantum field theories with Lagrangian description. More importantly, our study suggests that the reference state in the definition of complexity for a quantum field theory should be a simply state in the sense that there is no entanglement at all.

Interesting questions are proposed which can and should be studied in the soon future. For example how to understand the measure $d[\mu]$ and the function $Area(\mu)$ deeper on field theory side and how to realize an explicit example for the process of preparing quantum states using the $T\bar{T}$ deformation and hence understand the bulk entanglement wedge better\footnote{Some important progress in understanding the entanglement wedge as a dual of a boundary subregion density matrix has been achieved in \cite{Caputa:2017urj,Caputa:2017yrh}.}. Those questions are important for simulating a quantum field theory on a quantum computer.

\section*{Acknowledgement}
I appreciate Andreas Karch and Laurence Yaffe for useful discussions. This work was supported in part by a grant from the Simons Foundation (651440, AK). I am very grateful to my parents and recommenders.
\bibliographystyle{JHEP}
\bibliography{TTCV}
\end{document}